\documentclass{aastex62}

\usepackage{graphicx}
\usepackage{amsmath}
\usepackage{amssymb}
\usepackage{bm}
\usepackage{natbib}
\usepackage{color}
\usepackage[normalem]{ulem}

\usepackage{mymacros}

\received{\today}
\revised{\today}
\accepted{\today}
\submitjournal{ApJ}

\shorttitle{A Model for CME Energy Buildup and Eruption Onset}
\shortauthors{Dahlin et al.}

\begin{document}

\title{A Model for Energy Buildup and Eruption Onset in Coronal Mass Ejections}

\correspondingauthor{J.\ T.\ Dahlin}
\email{joel.t.dahlin@nasa.gov}

\author{J.\ T.\ Dahlin}
\altaffiliation{NASA Postdoctoral Program Fellow} 


\author{S.\ K.\ Antiochos}

\author{C.\ R.\ DeVore}
\affil{Heliophysics Science Division, NASA Goddard Space Flight Center, Greenbelt, Maryland 20771}



\begin{abstract}

  Coronal mass ejections (CMEs) and eruptive flares (EFs) are the most
  energetic explosions in the solar system. Their underlying origin is
  the free energy that builds up slowly in the sheared magnetic field
  of a filament channel. We report the first end-to-end numerical
  simulation of a CME/EF, from zero-free-energy initial state through
  filament-channel formation to violent eruption, driven solely by the
  magnetic-helicity condensation process. Helicity is the topological
  measure of linkages between magnetic flux systems, and is conserved
  in the corona, building up inexorably until it is ejected into
  interplanetary space. Numerous investigations have demonstrated that
  helicity injected by small-scale vortical motions, such as those
  observed in the photosphere, undergoes an inverse cascade from small
  scales to large, ``condensing'' at magnetic-polarity boundaries. Our
  new results verify that this process forms a filament channel within
  a compact bipolar region embedded in a background dipole field, and
  show for the first time that a fast CME eventually occurs via the
  magnetic-breakout mechanism. We further show that
  the trigger for explosive eruption is reconnection onset in the
  flare current sheet that develops above the polarity inversion line:
  this reconnection forms flare loops below the sheet and a CME flux
  rope above, and initiates high-speed outward flow of the CME. Our
  findings have important implications for magnetic self-organization
  and explosive behavior in solar and other astrophysical plasmas, and
  for understanding and predicting explosive solar activity.
\end{abstract}



\section{Introduction} \label{sec:intro}

Coronal mass ejections (CMEs) are spectacular manifestations of solar
activity in which enormous quantities of plasma and magnetic flux are
expelled from the solar atmosphere into interplanetary space. Although
they have been studied for decades, the mechanisms governing the
gradual energy buildup and explosive release that characterize all
CMEs remain subjects of intense investigation, due primarily to the
lack of accurate quantitative measurements of the magnetic field in
the corona. The many studies, however, have confirmed that all CMEs
and flares are
associated with solar filament channels \citep{Gaizauskas98,Martin98},
which consist of highly sheared magnetic field localized around
polarity inversion lines (PILs) of the radial component of the
photospheric magnetic field. This sheared flux is believed to provide
the free energy that powers the events. Understanding the energy
buildup mechanism leading to explosive eruption, therefore, comes down
to understanding how filament channels form.
\par

Although filaments form over all PILs, in quiet Sun as well as active
regions, and have been observed for many decades, their origin is
still under debate.  It is well known, however, that these structures
represent a balance of forces between the outward magnetic pressure of
the low-lying, strongly sheared flux and the inward magnetic tension
of the overlying, mostly unsheared flux.  CMEs are believed to be due
to the catastrophic disruption of this force balance inherent in all
filament channels \citep[see][and references therein]{Forbes06}. As a
result of the disruption, the free magnetic energy stored in the
corona is rapidly converted into bulk kinetic and thermal energy of
the plasma and nonthermal accelerated particles. A twisted flux rope
is ejected into the heliosphere, and hot X-ray emitting loops form
\textit{via} flare reconnection below the ejection. A central point of
controversy is the issue of cause and effect: is the flux rope formed
prior to the eruption, and does its ideal destabilization and
attendant outward motion induce the flare reconnection as a secondary
phenomenon \citep[e.g.][]{Fan01,Linker03b,Kliem06}? Or is reconnection
the primary actor in both the creation and fast expulsion of the flux
rope \citep[as in][]{Antiochos98,Antiochos99,Moore01,DeVore08,Karpen12,Wyper18}?
The answers to these questions are fundamental for understanding the
mechanism for explosive solar eruptions across all scales
\citep[][and references therein]{Wyper17}.
\par

The recently developed theory of helicity condensation
\citep{Antiochos13} proposes an answer to the question of filament
channel formation or, equivalently, the mechanism for energy buildup.
This theory posits that the photospheric convective motions
responsible for coronal heating continually inject magnetic
twist/stress into the corona at small scales, and that this twist
subsequently is transported by magnetic reconnection to collect at the
boundaries of coronal flux systems (PILs) at large scales. Magnetic
helicity is a topological measure of flux linkages in magnetized
plasmas \citep{Berger84a,Finn85}.  It is rigorously conserved
under ideal motion and is well preserved during reconnection in highly
conducting environments such as the corona \citep{Berger84b}. This
conservation property implies that the helicity associated with the
injected small-scale twist/stress must accumulate in the corona
because it cannot be dissipated away locally. Detailed numerical
simulations
\citep{Zhao15,Knizhnik15,Knizhnik17a,Knizhnik17b,Knizhnik18} have
confirmed the essential features of the process described by
\citet{Antiochos13} and demonstrated that the helicity indeed
``condenses'' at PILs to form stable magnetic structures whose
properties are consistent with those of observed filament channels.
A statistically averaged approximation to the full helicity-condensation 
process has been used by \citet{Mackay14,Mackay18} to investigate the 
formation of filament channels across the full Sun and over multiple solar 
rotation periods.
\par

The previously cited detailed numerical simulations of helicity injection,
transport, and condensation were performed for a closed Cartesian domain and
a purely bipolar magnetic field. Eruption was not possible in these
cases, because the upper boundary imparted an artificial confinement. These
studies, therefore, could not address the issue of the mechanism for
force-balance disruption, i.e., eruption onset. In this article, we
report the first numerical simulation of helicity condensation in a
fully 3D spherical system that realistically captures the structure
and dynamics of actual CMEs. The domain extends out to 30 solar radii,
so that the upper boundary has no effect on eruption
onset. Furthermore, we assume for the magnetic topology a localized
bipole, representative of a simple active region, embedded in a
background global dipole field. As described in detail below, we find that this simulation captures, for the first time, the self-consistent
filament channel formation via helicity condensation together with the explosive eruption that characterizes CMEs. Furthermore, we find that the
eruption is driven by magnetic reconnection, rather than by an ideal
instability.
\par

\section{Numerical Model} \label{sec:model}

\begin{figure}
\plotone{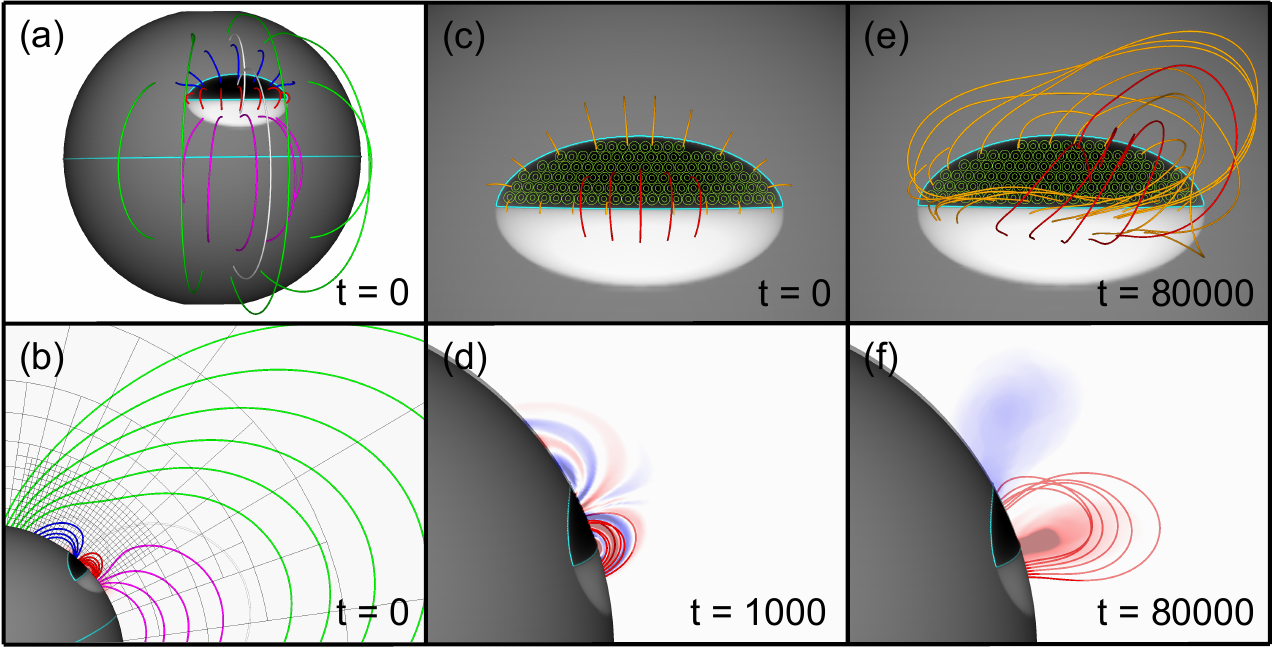}
\caption{Initial configuration (a,b) and filament-channel formation (c-f) at the indicated times $t$. Black (white) surface shading indicates minority (majority) magnetic polarity; cyan curves are the surface PILs. Green, blue, red, and magenta field lines illustrate different flux systems in the multipolar topology. Green annuli (c,e) are contours of constant $|\mathbf{v}|$ in the vortical flows, and orange field lines represent low-lying sheared flux above the PIL. Color shading in the $\phi = 0$ plane (d,f) contours the azimuthal field component, $B_\phi$, saturated at (d) $\pm 2$ G and (f) $\pm 30$ G.
An animation of panels (c,e) and (d,f) is available. The video starts at $t=0$ and ends at $t=80000$. Its duration is 5 seconds.}
\label{fig:config}
\end{figure}

Our numerical simulation was performed with the
Adaptively Refined Magnetohydrodynamics Solver
\citep[ARMS;][]{DeVore08}, which has been used to model both
CMEs/eruptive flares \citep[see
also][]{Lynch08,Lynch09,Karpen12,Masson13,Lynch16} and the formation
of filament channels
\citep{Zhao15,Knizhnik15,Knizhnik17a,Knizhnik17b,Knizhnik18}. The
magnetic field configuration shown in Figure \ref{fig:config}a,b
consists of a potential bipolar active region centered at 22.5$^\circ$ N
latitude with peak field strength $|B_r| \approx 50$ G \citep[adapted
from][]{DeVore08} embedded in a background dipole field of strength
$10$ G. The resulting topology is that of the well-known embedded
biopole with its separatrix dome and a pair of spine
lines emanating from a 3D null point \citep{Antiochos90,Lau90,Priest96}.
A region of maximally refined grid enclosed the coronal null
point, the entire separatrix dome below the null, and a substantial
additional volume above the null, in order to resolve the small-scale
reconnection dynamics of the helicity-condensation process (Fig.\
\ref{fig:config}b). The initial atmosphere was a spherically symmetric
hydrostatic equilibrium with an inverse-$r$ temperature profile
at a base temperature $T_s = 2 \times 10^6$ K and pressure
$P_s = 4 \times 10^{-1}$ dyn cm$^{-2}$ \citep[adapted
from][]{Karpen12}. We solved the ideal MHD equations with an adiabatic
temperature equation \citep[as in][]{DeVore08,Karpen12}. The domain
extents were $r \in [1R_s, 30R_s]$, $\theta \in [\pi/16, 15\pi/16]$,
and $\phi \in [-\pi,+\pi]$.
\par

This system was energized by motions at the bottom photospheric
boundary consisting of 107 close-packed, vortical, $B_r$-conserving
cellular flows within the black-shaded, minority-polarity region (see
Fig.\ \ref{fig:config}c). The flows were subsonic and sub-Alfv\'enic,
attaining a maximum speed $|V_\perp| \approx 50$ km s$^{-1}$ after an
initial sinusoidal ramp-up interval $1,000$ s in duration. Note that
the flows injected both free energy and magnetic helicity throughout
the coronal volume interior to the separatrix, i.e., on all field
lines that root at both ends within the northern hemisphere (red lines
in Fig.\ \ref{fig:config}b).  Everywhere
on the solar surface outside of the minority polarity, the magnetic
field was line-tied at rest ($\mathbf{V_\perp} = 0$) at all times
during the simulation. Hence, any changes in the connectivity of field
lines outside of the initial separatrix dome occurred solely due to
reconnection in the corona. Both reconnection and convection
contributed to connectivity changes inside the dome. Except for the
localized dynamics due to reconnection, the pre-eruption evolution was
approximately quasi-static, as in the real corona, because the average
photospheric driving speed ($\approx 25$ km s$^{-1}$) was much smaller than the typical Alfv\'en speed in the bipolar active region ($\approx 1,000$ km s$^{-1}$).
\par

\section{Results}\label{sec:results}

In this section, we present the results of our simulation, tracing the
evolution of the system through the self-consistent formation of the
filament channel (\S\ref{subsec:filament}) to its eventual explosive
eruption in a CME (\S\ref{subsec:eruption}). We then examine in detail
the timings of impulsive flare-reconnection onset and kinetic- and
magnetic-energy changes (\S\ref{subsec:energetics}) and the
topological evolution of the magnetic field (\S\ref{subsec:topology}).

\subsection{Energy Buildup}\label{subsec:filament}

The formation of the filament channel is shown in Figure
\ref{fig:config} at times early ($t = 1,000$ s, when the vortical
flows attain their peak speed) and late in the evolution, ($t = 80,000$ s, not long
before onset of eruption). Panels d,f show the
shear component of the magnetic field, $B_\phi$, color-shaded against
the sky at the central meridian ($\phi = 0$). The alternating red and
blue contours in Figure \ref{fig:config}d mark the weakly twisted
structures generated initially by the individual vortical cells. Over
time, these alternating bands of shear field strengthen and reconnect
with their neighbors, transporting the shear flux toward the outer
boundary of the minority-polarity region, i.e., the PIL. Figure
\ref{fig:config}f shows the resulting concentration into a localized
region above the PIL, consistent with previous modeling of helicity
condensation \citep[e.g., cf.\ Fig.\ 4 in][]{Knizhnik15}. The
three-dimensional filament channel penetrates the meridional plane
twice, yielding two oppositely signed regions of $B_\phi$. Viewed from
above in Figure \ref{fig:config}e, the channel is seen to be
quasi-circular and to envelope the whole PIL. It consists of an arcade
of highly sheared, low-lying flux (orange) beneath weakly sheared,
high-lying flux (red, above the southern segment of the PIL).
\par

A key point is that the simple driver of our system consists solely of
small-scale, compact, close-packed, vortical flows. If the system
evolution were purely ideal, these flows would never produce the
global-scale filament channel evident in Figure \ref{fig:config}. The
combination of the flows with helicity-conserving coronal magnetic
reconnection, however, produces a very highly sheared filament channel
whose properties are consistent with solar observations. As Figure
\ref{fig:config}f indicates, the resulting self-organized structure
consists of a balance of forces between an outward magnetic pressure
exerted by the out-of-plane field component and an inward magnetic
tension exerted by the stretched in-plane field components. We discuss
below how this force balance is disrupted.
\par

\subsection{Eruption}\label{subsec:eruption}

\begin{figure}
\plotone{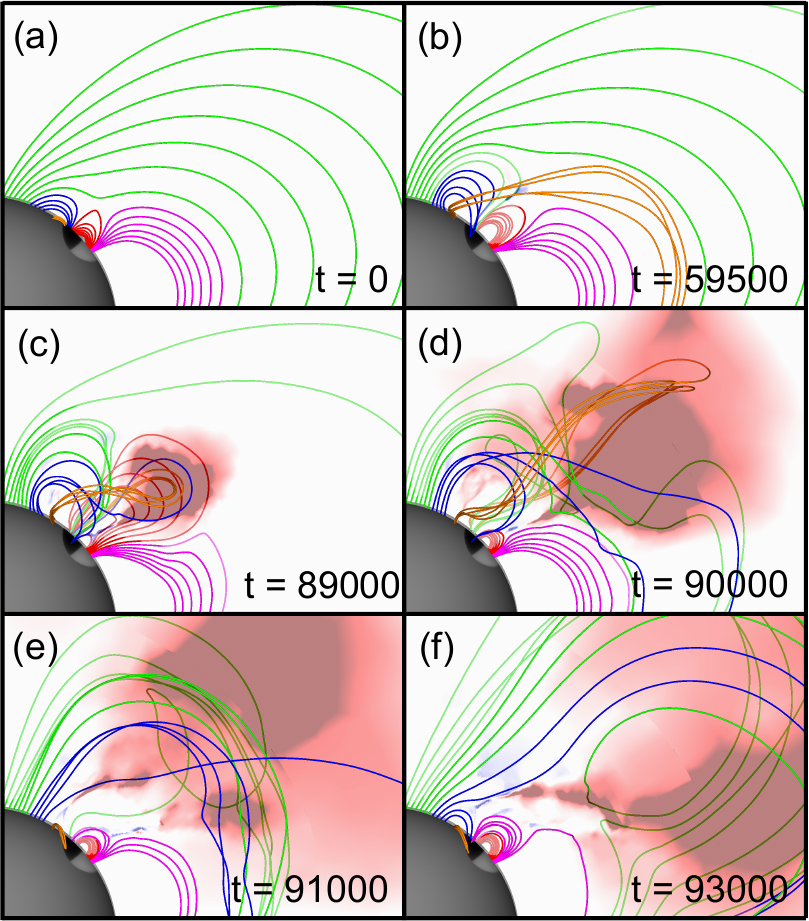}
\caption{Filament-channel eruption. Surface shading and magnetic field-line coloring are the same as in Figure \ref{fig:config}. Red (blue) shading in the semi-transparent plane at $\phi = 0$ shows outward (inward) radial flow saturated at $\pm 1,000$ km s${^{-1}}$.
An animation is available. The video starts at $t=86600$ and ends at $t=99100$. Its duration is 15 seconds.}
\label{fig:eruption}
\end{figure}

The buildup of the filament channel shown in Figure \ref{fig:config}
strongly perturbs the overlying magnetic field
and sets the stage for the later eruption. As shown in Figure
\ref{fig:eruption}, between times $t = 0$ (a) and $t = 59,500$ s (b),
the expansion of the separatrix dome moves the coronal magnetic null
point outward and slightly equatorward. This motion also compresses the null to
form a current sheet, which allows slow magnetic-breakout reconnection
to begin there. Evidence for this can be seen in Figure
\ref{fig:eruption}b: the two innermost green field lines that
initially connected into the southern hemisphere now connect into the
minority-polarity region in the north. This has removed flux that
overlay the separatrix dome and the southern portion of the PIL by
transferring it to overlie the northern portion of the PIL, weakening
the restraining tension force above the dome and enabling further
expansion. Additional evidence of breakout reconnection is shown by
the orange field lines, which now connect to the southern hemisphere
rather than into the minority-polarity region in the north as they did
previously.
\par

By time $t = 89,000$ s in Figure \ref{fig:eruption}c, the eruption is
well underway. All but two of the green field lines now connect into
the minority-polarity region and no longer overlie the southern
portion of the PIL, where a section of the strongly sheared filament
channel (bundle of orange field lines) is erupting outward at high
speed, $v_r > 1,000$ km s$^{-1}$ (dark red shading). At time
$t = 90,000$ s in Figure \ref{fig:eruption}d, the ejected sheared flux
has more than doubled its height above the surface, and reconnection
has commenced below where the fluxes from each side have closed back
together in the wake of the CME. This is evidenced by the green field
lines that have reestablished their connections to the southern
hemisphere at this time. Paired upflows (red) and downflows (blue)
highlight the flare reconnection jets beneath the high-speed ejecta.
\par

The flare reconnection not only restores all of the green field lines
to their original connections to the southern hemisphere, as seen in
Figure \ref{fig:eruption}e, but also creates entirely new connections
of blue field lines to the southern hemisphere, whereas previously
they connected only into the minority polarity in the north. This
process weakens the field overlying the northern portion of the PIL,
so now that segment of the sheared filament channel also erupts at
high speed in a ``sympathetic'' ejection. (The orange sheared field
lines that are drawn participate only in the first eruption, not the
second.) This second eruption is deflected equatorward by the
overlying field, as can be seen at time $t = 93,000$ s in Figure
\ref{fig:eruption}f. The flare reconnection jets below the second CME
also are clearly visible in this image. Therefore, although the
eruptive event begins in the filament channel above the southern
portion of the PIL, it eventually encompasses the entire channel and
PIL encircling the minority-polarity region.
\par




\subsection{Energetics and Timings}\label{subsec:energetics}

\begin{figure}
\plotone{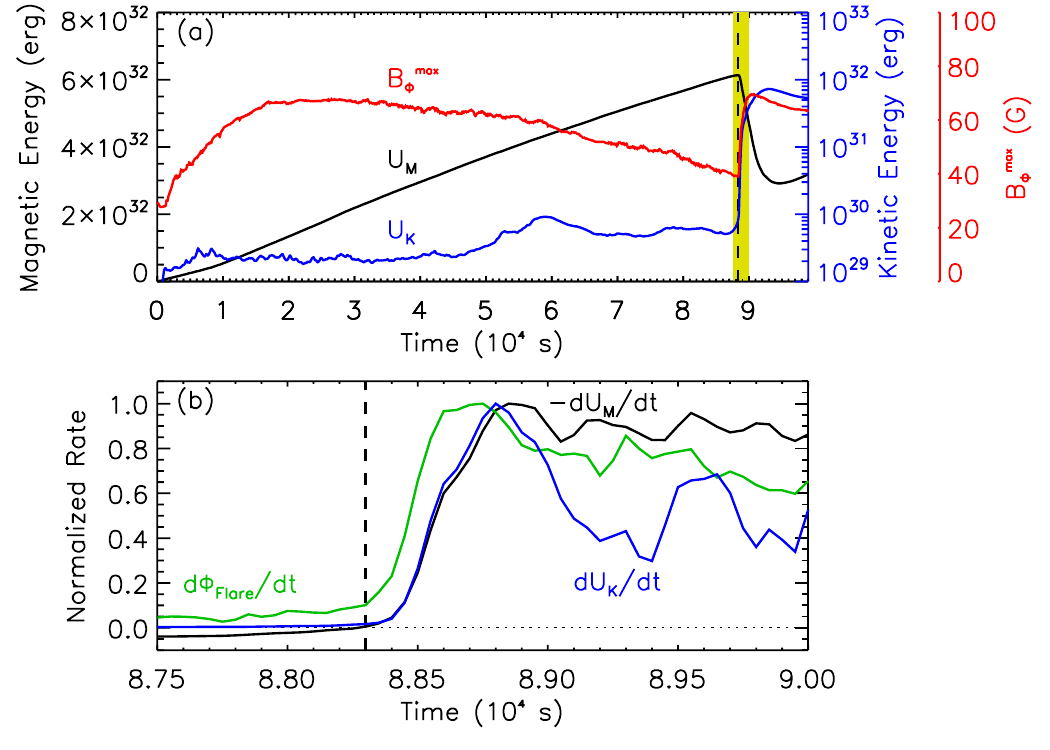}
\caption{Global energetics. (a) Time histories of total magnetic ($U_M$) and kinetic ($U_K$) energies and of maximum longitudinal field strength ($B_\phi^{\rm max}$); the yellow-shaded region is the temporal interval expanded in the bottom panel. (b) Rates of change of total magnetic and kinetic energies and of flare-reconnected flux, normalized to their respective peak values ($dU_M/dt=1.1 \times 10^{29}$ erg s $^{-1}$; $dU_K/dt=4.4 \times 10^{28}$ erg s $^{-1}$; $d\Phi_{\text{Flare}}/dt=5.5 \times 10^{18}$ Mx s$^{-1}$), around the time of eruption onset marked by the dashed vertical line in both panels.}
\label{fig:energetics}
\end{figure}

To clarify the mechanism for eruption onset, we display the global
energetics and other properties of our simulation in Figure
\ref{fig:energetics}. Full time histories of the total,
volume-integrated magnetic ($U_M$; black curve) and kinetic ($U_K$;
blue curve) energies are shown in Figure
\ref{fig:energetics}a. Filament-channel formation occurs up to time
$t \approx 88,300$ s, marked by the vertical dashed line in the plot,
when the eruption begins. During the formation phase, the vortical
surface driving builds up the free magnetic energy at a nearly constant rate,
until this energy eventually reaches $U_M \approx 6 \times 10^{32}$
erg, which is about 20\% of the initial potential-field energy. After a
brief startup phase, the kinetic energy $U_K$ exhibits two extended
intervals of quasi-steady values: $\approx 2 \times 10^{29}$ erg
early, and $\approx 5 \times 10^{29}$ erg later, with a transition
occurring between $t \approx 50,000$ s and $t \approx 60,000$ s. This
transition coincides with the onset of breakout reconnection in the
corona, as discussed and illustrated in the preceding \S
\ref{subsec:eruption} and Figure \ref{fig:eruption}. The early
behavior is associated primarily with the imposed vortical flows and
the pervasive reconnection outflows that follow from the
helicity-condensation process; the late evolution has an additional
kinetic-energy contribution due to the slow outward expansion of the
filament channel and the overlying separatrix dome, along with some
reconnection outflows near the null point. All of this flow energy
remains small, amounting to only about 0.1\% of the stored magnetic
free energy, during the buildup phase.
\par

Also shown in Figure \ref{fig:energetics}a is the maximum longitudinal
field strength ($B_\phi^{\rm max}$; red curve) within the
computational domain. This quantity provides an approximate measure of
the maximum magnitude of the shear field in the filament channel. The
initial value in the potential field is about 30 G. As the vortical
surface driving twists the magnetic field lines within the
minority-polarity region, the maximum strength increases, peaking at
about 70 G at time $t \approx 20,000$ s. Thereafter, although the
amount of shear flux continues to increase as the helicity condenses
to build the filament channel, the upward expansion increases the
cross-sectional area of the channel fast enough that the strength of
the shear field declines. This result is expected from the classic
arguments of the Aly-Sturrock free-energy limit (Aly 1991, Sturrock 1991). We note
that there is a subtle, but perceptible, steepening of the slope of
the $B_\phi^{\rm max}$ curve simultaneous with the increase in kinetic
energy, associated with the onset of slow breakout reconnection as
noted above.

The initiation of the filament-channel eruption at $t \approx 88,300$
s (dashed line) in Figure \ref{fig:energetics}a is accompanied by
steep changes in all three quantities. The kinetic energy rises by
more than two orders of magnitude, reaching a large fraction (about
10\%) of the total stored magnetic free energy. This indicates a
sudden transition from very low-speed flows to very high-speed flows,
whose average Alfv\'en Mach number is about 0.2. The maximum
longitudinal field strength nearly doubles very quickly as
the shear flux below the eruption is compressed by the downward
contracting flare loops (e.g., Zucarello et al 2017). Energy is
extracted from the magnetic field to power the high-speed flows and
the local $B_\phi$ compression, as indicated by the approximately 50\%
drop in magnetic energy during the eruption.


In order to understand in more detail the transition from channel
formation to eruption, we expanded the small interval indicated by the
yellow shading in Figure \ref{fig:energetics}a for display in Figure
\ref{fig:energetics}b. There we show the rates of change of the total
magnetic ($dU_M/dt$; black) and kinetic ($dU_K/dt$; blue) energies,
along with that of the total flare-reconnected flux
$d\Phi_{\rm flare}/dt$; green), which we measured by tracking
discontinuous changes in the lengths of magnetic field lines rooted
in the minority-polarity region. This plot shows that strong flare
reconnection begins first, and is followed by initiation of the
rise in kinetic energy simultaneous with the decline in magnetic
energy. This sequence demonstrates conclusively that reconnection
plays the primary role in initiating the explosive eruption of the
filament channel, and is fully consistent with previous work on
magnetic breakout for both CMEs/EFs \citep{Karpen12} and coronal jets
\citep{Wyper17,Wyper18}.

\subsection{Topological Evolution}\label{subsec:topology}

\begin{figure}
\plotone{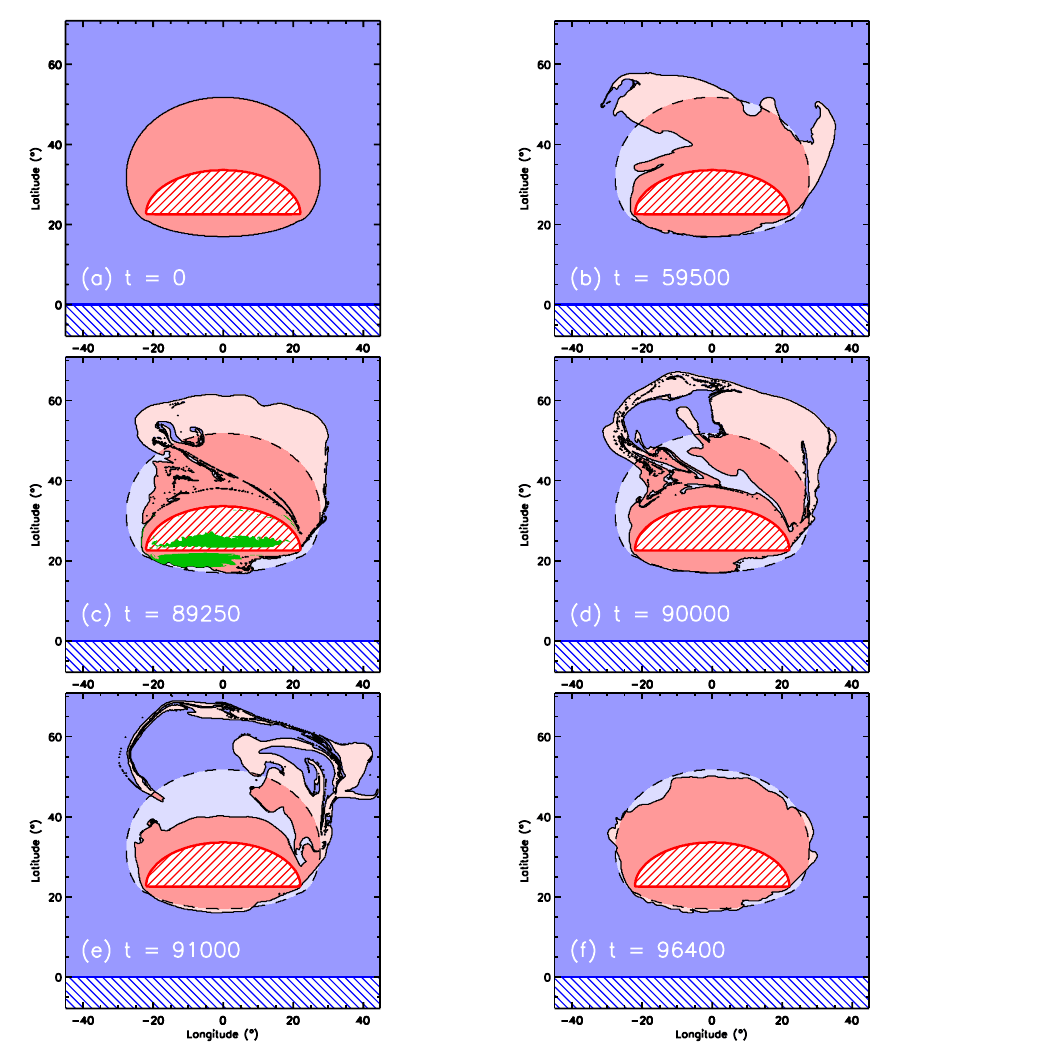}
\caption{Connectivity maps at \textbf{six} steps in the evolution. Dark red/blue shading with diagonal lines indicates minority-polarity flux in the northern hemisphere and all flux in the southern hemisphere, respectively. Medium red/blue shading indicates flux that closes to the northern/southern hemisphere. Light red/blue shading indicates flux that closes to the northern/southern hemisphere that initially connected to the opposite hemisphere. Solid black lines mark the intersection of the separatrix surface with the photosphere; dashed black lines mark the initial location of this intersection. Green shading in (c) indicates flux that has undergone flare reconnection.
An animation is available. The video starts at $t=0$ and ends at $t=96400$. Its duration is 15 seconds.}
\label{fig:topology}
\end{figure}

The reconnection dynamics that accompany the filament-channel
formation and eruption are illuminated by examining the topological
evolution of the system in Figure \ref{fig:topology}. Figure
\ref{fig:topology}a depicts the initial magnetic connectivity. The
solid-shaded medium blue region in the north maps to the southern
hemisphere (diagonal dark blue shading), while the solid-shaded medium
red region in the north maps to the minority-polarity flux in the
active region (diagonal dark red shading). The black contour is the
intersection of the separatrix dome, which encloses all of the flux
connecting to the minority-polarity region, with the solar
surface. All field lines outside of the minority-polarity region are
line-tied and fixed ($\mathbf{v}_\perp = 0$), and only those lines are
used to compute the connectivity maps. Hence, all changes in the
connectivity shown in Figure \ref{fig:topology} occur due to
reconnection.

As the minority-polarity flux is twisted by the vortical motions and
then accumulates concentrated shear via helicity condensation,
magnetic pressure builds up within the separatrix dome and deforms the
coronal null to form a current sheet. Over time, the resulting
reconnection changes the connectivity of the flux at the solar
surface. By time $t = 59,000$ s, some of the majority-polarity flux
that initially closed within the separatrix dome (medium red, Fig.\
\ref{fig:topology}a) has reconnected and now closes to the southern
hemisphere (light blue, Fig.\ \ref{fig:topology}b). Correspondingly,
an equal amount of the majority-polarity flux that initially closed to
the southern hemisphere (medium blue, Fig.\ \ref{fig:topology}a) now
closes to the minority-polarity region (light red, Fig.\
\ref{fig:topology}b). Thus, the breakout reconnection has the effect
of transferring the flux that overlies the southern segment of the PIL
(the straight portion of the dark red minority-polarity boundary)
so that it overlies the northern segment (the curved portion of that boundary).

This process continues and expands the area of changed connectivity
until the flux overlying the southern segment of the PIL weakens
sufficiently to allow eruption to occur.  Figure \ref{fig:topology}c
shows the topology at $t = 89,250$ s, shortly after flare-reconnection
onset and initiation of the eruption above the southern segment of the
PIL (Fig.\ \ref{fig:eruption}c). Additional flux has been removed from
above the southern segment of the PIL (see in particular the light
blue regions in the southern half of the active region, where a small
area corresponds to a comparatively large amount of flux due to the high field strength). The flux
that has undergone closed-closed flare reconnection within the
separatrix dome is shown in green. This region corresponds to the
classic flare ribbons. It maps out a region along the southern segment
of the PIL and, due to the shear concentrated there, the kernel of
reconnected flux south of the PIL is displaced to the east. We note
also that the separatrix boundary has become highly fragmented and
fractal in character, possessing structure down to the
grid scale.

As the eruption proceeds, the connectivity continues to evolve
in structure and complexity. Subsequent maps are shown at times
$t = 90,000$ s and $91,000$ s in Figures \ref{fig:topology}d,e for
comparison with the preceding Figures \ref{fig:eruption}d,e. As the
flare reconnection progressively processes flux in the green and
blue flux systems drawn in Figure \ref{fig:eruption} at these times,
the connectivity of the field lines switches from the
minority-polarity region in the northern hemisphere to the southern
hemisphere. This change is indicated by the new light blue regions
adjacent to the northern separatrix boundary in Figure
\ref{fig:topology}, especially at the later time (Fig.\
\ref{fig:topology}e). The strong reduction in the flux overlying the
northern segment of the PIL enables that section of the filament
channel to erupt, as we noted above in \S3.3. In the wake of that
secondary ejection, flare reconnection above the northern segment of
the PIL then acts to restore the connections of the blue field lines
from the southern hemisphere back to the north, as shown in Figures
\ref{fig:eruption}e,f. This is reflected in the change in the
connectivity-map colors from light blue to medium red in Figures
\ref{fig:topology}e,f.

After the conclusion of the two episodes of CME ejection and flare
reconnection, above first the southern and then the northern segments
of the PIL, the system contains subcritical amounts of helicity and
magnetic free energy and relaxes toward a new equilibrium state. The
connectivity at the end of the simulation is shown in Figure
\ref{fig:topology}f. The separatrix dome has taken on a configuration
that rather closely resembles the initial state, and all of the
connectivity changes at this time are localized near the original
separatrix boundary. Ongoing helicity condensation now can rebuild the
free energy and helicity in the channel until it once again attains
the critical condition for a new eruption.

\section{Discussion\label{sec:discussion}}

This article presents the first calculations of a CME powered by
self-consistent energy buildup via helicity condensation. We have
shown that a coronal field configuration driven by helicity-injecting
flows leads to a stressed magnetic structure consistent with
observations and that this structure eventually erupts
explosively. Breakout reconnection above destabilizes the system, and the
onset of flare reconnection below produces explosive CME acceleration and
rapid magnetic energy release, generating and ejecting a flux
rope.

The eruption phases discussed in \S\ref{subsec:energetics} are largely
consistent with the results of prior high-resolution 2.5D studies
\citep{Karpen12}. However, several new aspects arise due to the
three-dimensional topology. The PIL in these calculations is
semicircular, rather than an axisymmetric line. Viewed from the
mid-plane (as in Fig.\ \ref{fig:config}f), it can be seen that the two
locations for shear accumulation and magnetic pressure stress the
null point in orthogonal directions, inhibiting its deformation into a
current sheet. This allows more energy to build up prior to eruption,
explaining why the maximum free energy ($\approx 21\%$) is
substantially larger than in 2.5D simulations ($\approx 11\%$). The entire filament
channel is ejected; the southern half erupts first and the northern
half follows ``sympathetically,'' as is frequently observed.

Another consequence of the three-dimensional configuration is a
rotational asymmetry in the breakout reconnection (see in particular
Fig.\ \ref{fig:topology}). Due to the large-scale twist in the field
inside the separatrix dome, when the flux above the southern half of
the PIL undergoes breakout reconnection, the resulting reconfigured flux
is positioned on the east of the minority polarity. The overlying field
accumulates more shear compared to prior studies of the helicity
condensation process \citep[cf.\ Fig.\ 3 in][]{Knizhnik17b}. This
may be because only six cellular flows span the narrowest part of
the minority polarity; on the other hand, observations of coronal cells
indicate that the magnetic-shear region extends a substantial
distance from the filament channel \citep{Sheeley13}.

The primary conclusion from our calculation is that a broad variety of solar
features, both pre- and post-eruption, follow directly from a turbulent-like
inverse cascade of magnetic helicity in a simple multipolar topology.
These ingredients are ubiquitous in the solar corona. In this
Study, we assumed a set of uniform rotational surface flows for
simplicity; however, the particular form of the driving
should not be important, so long as there is a net injection of
helicity and sufficiently complex flows to drive the pervasive
reconnection required for the cascade. Net helicity may be injected
into the corona in other ways, such as through both flux emergence and
cancellation, as well as by large-scale flows as in active-region
rotation. Regardless of the source of the injected helicity,
stochastic coronal reconnection will concentrate
structure at PILs and inevitably lead to explosive eruption. 

Our work was supported by NASA's LWS, H-SR, H-ISFM, and HEC research
programs. Joel Dahlin acknowledges support from the NASA LWS Jack Eddy 
Fellowship administered by the University Corporation for Atmospheric Research.
Joel Dahlin's research was also supported by an appointment to the NASA Postdoctoral
Program at the NASA Goddard Space Flight Center, administered by Universities Space 
Research Association under contract with NASA.
We also acknowledge use of Peter Wyper's magnetic-topology tools to generate Figure \ref{fig:topology}.



\newpage

\bibliography{bibliography-final.bib}




\end{document}